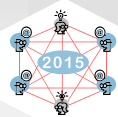

# Improvements in OMNeT++/INET Real-Time Scheduler for Emulation Mode

Artur Austregesilo Scussel

Instituto Tecnológico de Aeronáutica - ITA
São José dos Campos São Paulo, Brasil
email: arturscussel@gmail.com

Georg Panholzer, Christof Brandauer, Ferdinand von Tüllenburg

Advanced Networking Center
Salzburg Research Forschungsgesellschaft mbH
Salzburg, Austria
email: {firstname.lastname}@salzburgresearch.at

*Abstract*—In this paper the performance of INET's emulation mode is evaluated. In particular, the focus of the study is on the precision of the delay emulation. It is shown, that this precision is low in INET 2.6 (respectively a later version provided in the integration branch that fixed the crashes of 2.6 in emulation mode). Two errors in the implementation are identified and an alternative configuration for packet capturing is proposed. The performance tests are re-run with such a modified version of the real-time scheduler (which is now included in the recent INET 3.0 release) and it is shown that the responsiveness of the emulation mode and the precision of delay emulation are improved significantly. Finally, the negative impact of the modified capturing configuration is briefly analyzed. Packet loss in the capturing process has deteriorated but in fact is has already plagued the emulation mode of previous implementations and this topic clearly demands for further studies.

*Keywords–OMNeT++; INET framework; Emulation; Real-time simulation*

## I. INTRODUCTION

The work presented here was carried out in the course of an ongoing research project where inter-dependencies between the quality of an electrical energy network and its supporting communication network are analyzed. In the given scenario, power quality metrics are transmitted from a power system simulator to a real-world remote control application which computes new configuration parameters (e.g. setpoints) of the electrical network and send them back to the power system (simulation). The IEC 61850-based communication takes place via an emulated wide-area TCP/IP network.

For the study it is necessary to impair the communication quality (delay, loss, reordering etc.). Typically this is done through a software- (e.g. netem [1] or DummyNet [2]) or hardware-based WAN emulator. However, the next step of the study requires that the real-world communication flow will be multiplexed with many additional simulated flows as produced by simulated energy applications. Ideally, OMNeT++ could be used not just for the simulation but also for the emulation task by utilizing the emulation capabilities of the INET module [3]. It is therefore evaluated if the main requirement - a delay emulation precision in the order of a few milliseconds - can be met by INET's emulation mode. The test environment is *Host1*, a modern PC (Xeon E5-1650v3, 32GB RAM, Intel I350-T4 NIC) running Ubuntu Linux 14.04.

## II. PERFORMANCE EVALUATION

Several research works based on INET emulation have been published [4], [5], [6], [7], [8]. This however, is already some years back and when we started our work in December 2014 it was even difficult to get emulation mode up and running. The then current stable INET version 2.5.1 crashed upon the first packet arrival and the development version 2.99.0 didn't compile and had the same runtime problems once the compilation problems were sorted out. In fact, INET 2.1 (released in January 2013) was the last version where the emulation mode was running out of the box. It seems that INET emulation has not been widely used in the recent years.

Less than 1 day after we sent a bug report for version 2.5.1 to the OMNeT++ mailing list[1] a fix was provided in the integration branch by Zoltán Böjthe on December 12, 2014. Based on this INET version and OMNeT++ version 4.6 the evaluation as presented in this paper was conducted.

### A. Ping to local StandardHost

The first test is based on the `extClient` example that is shipped with INET. The simulation model contains a single StandardHost where 1 external interface (`ext0`) is configured with an IP address of 10.1.1.1. IP traffic targeted to this address is captured on interface `eth1` and injected into the simulation. The real-world `ping` application is used to generate traffic (and receive the replies from the simulated StandardHost). It is run on the machine where OMNeT++ is installed and simply pings the address 10.1.1.1. It thus generates ICMP echo requests and for each ICMP echo reply received it prints the measured round trip time (RTT). OMNeT++ is run from the command line using the `Cmdenv` interface.

In contrast to our expectations, the RTT ranges from approximately 1 ms up to 22 ms with an average of 12 ms. A boxplot of the measured ping RTTs is depicted in the left-hand side ("ping local") of Figure 2. The lower and upper hinges of the boxplot cover the first and third quantiles of the measured values. The whiskers extend to the minimum and maximum of the measured values, respectively.

If, in comparison, the emulation is not used and the ping targets the IP of the local `eth1` interface, the RTTs are always below 1 ms.

### B. Ping to router over emulated link

In the second test the simulation model contains 2 routers that are connected by a DatarateChannel with a datarate of 1 Gbps and a delay of 10 ms (in each direction). Additionally, each router has 1 external interface. They are connected to the simulation machine's real interfaces `eth1` and `eth2`, respectively. These interfaces are both connected to remote PCs as shown in Figure 1.

---

[1] https://groups.google.com/forum/#!forum/omnetpp

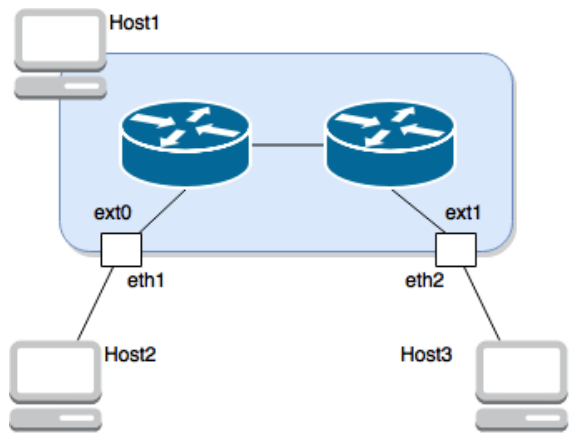

Figure 1. Setup for the performance tests with an emulated link.

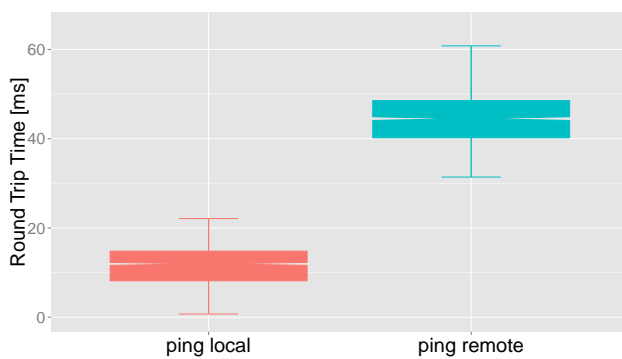

Figure 2. Boxplot for RTTs to local and remote nodes: original code.

Again, a simple ping test is run to obtain an approximate measure of the RTTs. Concretely, *Host2* pings *Host3* over *Host1*. To get a performance baseline first, *Host1* is configured for IP forwarding and OMNeT++ is not run. In this case, the RTTs are permanently below 1 ms. Then, IP forwarding is turned off and the packet forwarding on *Host1* is achieved via OMNeT++/INET. The emulated link between the two routers should increase the RTT by 20 ms and an overall RTT in the order of 21 ms is expected. A boxplot of the measured ping RTTs is depicted in the right-hand side ("ping remote") of Figure 2.

In both tests, the delay emulation is imprecise. The RTTs are significantly higher than what can reasonably be expected and exhibit strong variations. The precision was insufficient for the given requirements of the planned study and we thus analyzed whether (and how) the precision of delay emulation could be improved.

### III. ANALYSIS

In emulation mode events are not processed immediately one after the other as fast as possible but instead the processing of events is synchronized with the wall-clock time. This mode is implemented by INET's (soft) real-time scheduler (linklayer/ext/cSocketRTScheduler.cc).
Additionally, the scheduler enables the integration of real-world network traffic, i.e., real packets can be injected into the simulator and emitted from the simulator onto the real network.

In INET the capturing of packets is implemented via the pcap library [9] which provides a high-level interface to the kernel-level packet capturing facilities for several platforms. Packets are "sniffed" from a real network interface (e.g., eth0) and passed to a so called "external" interface of an OMNeT++ host (e.g., ext0) by inserting an event into the simulator's future event set (FES). An OMNeT++ Standard-Host/Router module already contains a configurable number of external interfaces. To emit packets to a real network, raw sockets are used.

INET's cSocketRTScheduler integrates simulation internal event processing with external packet events (that naturally occur simultaneously) in the following way: when the next simulation event is in the future and the simulation is thus idle (because the event processing must be synchronized with the wall-clock time), the real-time scheduler utilizes this idle time to check if new packets destined to the simulator have arrived on the host's network interface(s).

In search of performance bottlenecks of INET's emulation mode, we realized that this strategy was not implemented correctly. When the next event from the FES occurs at targetTime, the scheduler calls receiveUntil(targetTime). Internally, however, this function invokes receiveWithTimeout() which makes a blocking call to the pcap library with a fixed timeout of 10 ms (as defined hard-coded in PCAP_TIMEOUT). Thus, if the next event is, e. g., 3 ms in the future, the simulator would enter the blocking pcap function and stay there for 10 ms if no traffic destined for the simulator is observed on the host's interface(s). As a consequence, the processing of the next event is approximately 7 ms late. If this event results in emitting a packet on the real network, the packet is sent late.

Before we reported this problem (as we were not done with the overall analysis) it was obviously discovered elsewhere and an issue was filed[2] by András Vargas. A fix was provided by Rudolf Hornig and he additionally resolved the problem that Linux optimized code in the capturing process was *not* activated under Linux.

Another source of additional delays (and delay variations) in the INET emulation mode stems from the way the packet capturing framework is used. In the default mode, a packet is not immediately passed to the application when it is captured. Instead, packets are collected in the kernel until a timeout occurs or the receive buffer fills up (whichever occurs first) and are then passed to the application in one batch [10]. This strategy saves system resources (context switches, system calls) but introduces delays that may be undesirable for the real-time emulation mode. It can be disabled by invoking the pcap_set_immediate_mode before a capture handle is activated.

### IV. MODIFIED EMULATION CODE

To evaluate the impact of the above described issues the 3 changes (correct timeout computation; activation of Linux optimized capturing code; activation of the immediate mode via libpcap) were applied and the above tests were re-run. Indeed, the precision of the delay emulation is vastly improved as can be seen in Figure 3. The response times for local pings (ping local) are constantly below 1 ms and the ping RTTs measured over the emulated link (ping remote) are approximately 21 ms. In both cases, the variation of the observed delays is very small.

---
[2]https://github.com/inet-framework/inet/issues/120

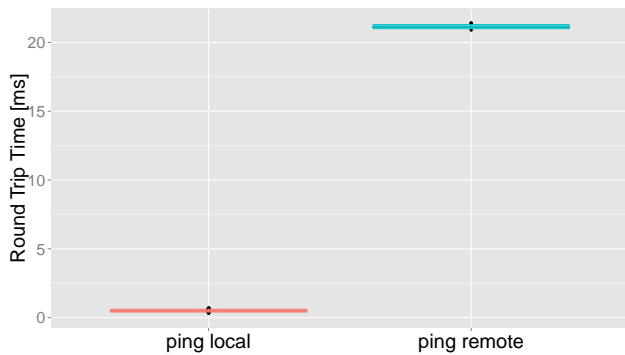

Figure 3. Boxplot for RTTs to local and remote nodes: modified code.

The tests were once again run against the recently released INET 3.0 version. It includes all the modifications described above. The results are exactly the same as shown in Figure 3.

The immediate mode in the capturing configuration doesn't come without a price. The potential drawback is the increased risk of packet loss in the capturing process since there is no buffer in immediate mode.

We ran several tests with various traffic rates and packet sizes on INET 3.0 with and without immediate mode. Indeed, packet loss in the capturing process could be observed in both versions and the loss was higher with immediate mode. As an example, a test with a generated traffic of 10 Mbit/s and a packet size of 100 Bytes (125 packets/s) yielded a capture loss of roughly 3.4 % (with immediate mode) and 1.6 % (without immediate mode). Other tests showed similar behaviour. These initial tests indicate that the immediate mode increases the packet loss in the capturing process, but packet loss can also happen without it. In any case, this is not acceptable as such a packet loss usually renders an emulation useless. Clearly, this topic requires further studies.

## V. CONCLUSIONS AND FUTURE WORK

In this paper we have studied the precision of delay emulation in the INET framework. The study was initiated after we faced unexpected delays with a high variance of delay values when using INET's network emulation mode. One key concern of this issue could be identified as a problem in the timeout computation within INET's real-time scheduler. This problem was fixed by OMNeT++/INET developers. However, this improvement is not sufficient to achieve delay values in a reasonable range. Further analyses of the pcap packet capturing mechanisms in the real-time scheduler has shown that using the pcap immediate mode is considerably reducing the delay.

Evaluations have shown, that with those code modifications both emulation responsiveness and precision of delay emulation could be significantly increased. The proposed code modifications were already integrated in version 3.0 of INET.

Unfortunately, our studies also unveiled packet loss in the packet capturing process ever for low bandwidth/packet rates. This issue is present with and without immediate mode. However, immediate mode increases packet loss due to disabled packet buffering at kernel space. However, this topic could only be covered superficially within this study and urgently needs further investigations. Possible approaches to face these issues would be a multi-threaded implementation of the packet capturing mechanism and/or usage of PF_RING [11], a new type of network socket that affords high-performance capturing under Linux and which is supported in current libpcap versions.

ACKNOWLEDGMENT

This work was carried out in the course of a project that is funded by the Climate & Energy Fund Austria within the "ENERGY MISSION AUSTRIA"' program.